\documentclass{article}
\usepackage{dcase2023,amsmath,url,times,booktabs,tabularx,amsmath,amssymb}
\usepackage{array,multirow}
\usepackage{arydshln}
\usepackage{multicol}
\usepackage{footnote}
\usepackage{float}
\usepackage{cite}
\usepackage{enumitem}
\usepackage{color,soul}

\usepackage[pdftex]{graphicx}

\newcolumntype{C}[1]{>{\centering\arraybackslash}p{#1}}
\newcolumntype{L}[1]{>{\raggedright\arraybackslash}p{#1}}
\newcolumntype{R}[1]{>{\raggedleft\arraybackslash}p{#1}}

\makeatother
\newcommand{\vect}[1]{{\mbox{\boldmath $#1$}}}
\newcommand{\T}[0]{\mathsf{T}}

\title{DESCRIPTION AND DISCUSSION ON DCASE 2024 CHALLENGE TASK 2: FIRST-SHOT UNSUPERVISED ANOMALOUS SOUND DETECTION FOR MACHINE CONDITION MONITORING}

\name{
Tomoya Nishida$^{1}$, Noboru Harada$^{2}$, Daisuke Niizumi$^{2}$, Davide Albertini$^{3}$, Roberto Sannino$^{3}$,
}
\secondlinename{
Simone Pradolini$^{3}$, Filippo Augusti$^{3}$, Keisuke Imoto$^{4}$, Kota Dohi$^{1}$, Harsh Purohit$^{1}$,
}
\thirdlinename{
Takashi Endo$^{1}$, and Yohei Kawaguchi$^{1}$
}
\address{
$^1$ Hitachi, Ltd., Japan, \url{tomoya.nishida.ax@hitachi.com}\\
$^2$ NTT Corporation, Japan, \url{noboru.harada.pv@hco.ntt.co.jp}\\
$^3$ STMicroelectronics, Italy, \url{}\\
$^4$ Doshisha University, Japan, \url{keisuke.imoto@ieee.org}\\
}

\begin{document}

\ninept
\maketitle

\begin{sloppy}

\begin{abstract}
We present the task description of the Detection and Classification of Acoustic Scenes and Events (DCASE) 2024 Challenge Task 2: ``First-shot unsupervised anomalous sound detection (ASD) for machine condition monitoring''. 
Continuing from last year's DCASE 2023 Challenge Task 2, we organize the task as a first-shot problem under domain generalization required settings. 
The main goal of the first-shot problem is to enable rapid deployment of ASD systems for new kinds of machines without the need for machine-specific hyperparameter tunings.
This problem setting was realized by (i) giving only one section for each machine type and (ii) having completely different machine types for the development and evaluation datasets.
For the DCASE 2024 Challenge Task 2, data of completely new machine types were newly collected and provided as the evaluation dataset.
In addition, attribute information such as the machine operation conditions were concealed for several machine types to mimic situations where such information are unavailable.
We will add challenge results and analysis of the submissions after the challenge submission deadline.
\end{abstract}


\section{Introduction}
\label{sec:intro}
Anomalous sound detection (ASD)~\cite{koizumi2017neyman, kawaguchi2017how, koizumi2019neyman, kawaguchi2019anomaly, koizumi2019batch, suefusa2020anomalous, purohit2020deep} is the task of identifying whether the sound emitted from a target machine is normal or anomalous.
This leads to automatic detection of mechanical failures, which is vital in the fourth industrial revolution with AI-based factory automation.
Using machine sounds for prompt detection of machine anomalies is useful for machine condition monitoring.

A major challenge concerning the application of ASD systems is that both the number and variety of anomalous samples can be inadequate in training.
In 2020, we held the first ASD task in Detection and Classification of Acoustic Scenes and Event (DCASE) Challenge 2020 Task 2~\cite{Koizumi2020dcase}; “unsupervised ASD” which aimed to detect unknown anomalous sounds using only normal sound samples as training data. 
Following this task, handling of domain shifts were additionally tackled in the DCASE Challenge 2021 Task 2~\cite{Kawaguchi2021} and 2022 Task 2~\cite{Dohi2022} for the wide spread application of ASD systems. 
Domain shifts are differences between the data in the source and target domains, which are caused by shifts in the operational conditions of the machine or environmental noise.
The DCASE Challenge 2021 Task 2~\cite{Kawaguchi2021}, “unsupervised ASD under domain shifted conditions” mainly focused on the use of domain adaptation techniques, whereas the DCASE Challenge 2022 Task 2~\cite{Dohi2022}, “unsupervised ASD applying domain generalization techniques” focused on the use of domain generalization techniques.

In the DCASE Challenge 2023 Task 2~\cite{Dohi2023DCASE}, "first-shot unsupervised ASD," real-world scenarios were explored even further as a "first-shot" ASD task.
In addition to the domain generalization problem setting, this task was designed as a scenario in which ASD needs to be performed on completely new machine types with data collected from only a single identification (ID) of that machine, without machine-specific hyperparameter tuning.
This scenario is typically encountered in many real-world situations where the rapid deployment of ASD systems is required and collecting a variety of training or test data is infeasible.
To realize this problem setting, the number of sections for each machine type was reduced to from multiple sections to one section, and the evaluation dataset consisted of completely new machine types unseen in the development dataset.
This setup prevented participants from performing handcrafted tunings or techniques that are difficult to implement in applications where rapid deployment of ASD systems are required.
For example, hyperparameter tuning for each machine type using the development dataset or training ASD systems with multiple sections of data has become infeasible.

To further deepen the techniques that are useful for this problem setting grounded on real-world scenarios, we designed the DCASE Challenge 2024 Task 2 "First-shot unsupervised anomalous sound detection for machine condition monitoring" by closely aligning to the problem setting established in the previous year.
The main modifications from DCASE 2023 Task 2 are that the evaluation dataset is updated with new machine types unseen in the previous DCASE ASD challenges, and that attribute information such as the machine operation conditions are concealed for several machine types.
The second modification concerns situations where such information is not unavailable, with the aim of expanding the range of applicable scenarios in real-world settings.

\section{First-shot Unsupervised Anomalous Sound Detection under Domain Shifted Conditions} 
\label{sec:uasd}
Let the $L$-sample time-domain observation $\vect{x} \in \mathbb{R}^L$ be an audio clip that includes sounds emitted from a machine.
The goal of the ASD task is to determine a given machine as either normal or anomalous by computing an anomaly score $\mathcal{A}_{\theta}(\vect{x})$ using an anomaly score calculator $\mathcal{A}$ with parameters $\theta$. 
The anomaly score is expected to be large for anomalous samples and small for normal samples. 
The input of $\mathcal{A}$ can be the audio clip $\vect{x}$ or $\vect{x}$ with additional information such as labels indicating the operation condition of the machine.
The machine is then determined to be anomalous when $\mathcal{A}_{\theta}(\vect{x})$ exceeds a pre-defined threshold $\phi$ as
\begin{equation}
\mbox{Decision} = \left\{
\begin{array}{ll}
\mbox{Anomaly} & (\mathcal{A}_{\theta}(\vect{x}) > \phi)\\
\mbox{Normal} & (\mbox{otherwise}).
\end{array}
\right.
\label{eq:det}
\end{equation}
The primary difficulty in this task is to train the anomaly score calculator with only normal sounds (unsupervised ASD). 
The DCASE 2020 Challenge Task 2~\cite{Koizumi2020dcase} was designed to address this issue and all the following tasks stand on this unsupervised ASD setting.

The domain-shift problem also needs to be solved for practical application of ASD.
Domain shifts are variations in conditions between training and testing phases that change the distribution of the observed sound data.
These shifts can arise from differences in operating speed, machine load, viscosity, heating temperature, environmental noise, microphone arrangement, and other factors.
Two domains, the \textbf{source domain} and the \textbf{target domain}, are defined: the former refers to the original condition with sufficient training data and the latter refers to another condition with only a few samples.
The 2021 Task 2 aimed to develop domain adaptation techniques, assuming the domain information (source/target) of each sample is known.
In contrast, the 2022 Task 2 aimed to develop domain generalization techniques.
Here, the domain information was assumed to be unknown in the test phase and anomalies from both domains had to be detected with a single threshold.
This year's task also inherits this domain generalization task setting.

Until 2022 Task 2, the provided data consisted of multiple sections for each machine type, with the development and evaluation datasets sharing the same machine types.
This allowed participants to train ASD models with handcrafted tunings for each machine type, e.g., hyperparameters or network architecture tunings using test data or data from similar machine types.
However, to further pursue the rapid development of ASD systems in real-world scenarios, solving ASD against completely novel machine types with only one section of data without handcrafted tunings are also highly important.
This is because in real-world scenarios, cunstomers may not possess multiple machines of the same machine type and collecting test data for handcrafted tuning may be infeasible.
This problem setting was named as the “first-shot problem”, and the 2024 Task 2 was organized based on this problem setting.
Specifically, the first-shot problem was realized by adding two features to the dataset:
(i) Completely different sets of machine types between the development and evaluation dataset and (ii) Only one section for each machine type.

While solving the first-shot problem under the domain generalization setting should be sufficient for many real-world applications, the results from the previous year suggested that there is still potential for further improvement in the solutions~\cite{Dohi2023DCASE}.
For example, less than half of the systems achieved better scores than the baseline system and even the top winning system could not beat the baseline system for two machines types in the evaluation dataset.
For this reason, we designed the DCASE Challenge 2024 Task 2, "First-shot unsupervised anomalous sound detection for machine condition monitoring" by closely aligning to the problem setting designed in the previous year.
The main modifications from 2023 Task 2 are that the evaluation dataset consists of newly recorded sounds of new machine types and that attribute information are concealed for several machine types.
By mostly following the same problem setting as in DCASE 2023 Task 2, the organizers aim to further deepen the techniques that are useful for first-shot ASD.

\section{Task Setup} 
\label{sec:task}

\subsection{Dataset} 
\label{sec:dataset}
The data for this task comprises three datasets: \textbf{development dataset}, \textbf{additional training dataset}, and \textbf{evaluation dataset}. 
The development dataset includes seven machine types, whereas the additional and evaluation dataset includes nine machine types, each having one section per machine type. 
\textbf{Machine type} means the type of machine such as fan, gearbox, bearing, etc.  \textbf{Section} is a subset or whole data within each machine type.
 
Each recording is a single-channel audio with a duration of 6 to 10 s and a sampling rate of 16 kHz. 
We mixed machine sounds recorded at laboratories with environmental noise recorded at factories and in the suburbs to create each sample in the dataset.
For the details of the recording procedure, please refer to the papers on ToyADMOS2~\cite{harada2021toyadmos2} and MIMII DG~\cite{Dohi2022}.

The \textbf{development dataset} consists of seven machine types (fan, gearbox, bearing, slide rail, valve, ToyCar, ToyTrain), and each machine type has one section that contains a complete set of the training and test data.
Each section provides 
(i) 990 normal clips from a source domain for training, 
(ii) 10 normal clips from a target domain for training, and
(iii) 100 normal clips and 100 anomalous clips from both domains for the test.
We provided domain information (source/target) in the test data for the convenience of participants. 
For four machine types (fan, bearing, valve, ToyCar), attributes that represent operational or environmental conditions are also provided in the file names and attribute csvs.
For the other three machine types, attributes are concealed.

The \textbf{additional training dataset} provides novel nine machine types (3D-printer, air compressor, brushless motor, hairdryer, hovering drone, robotic arm, scanner, toothbrush, ToyCircuit).
Each section consists of 
(i) 990 normal clips in a source domain for training and
(ii) 10 normal clips in a target domain for training.
For five machine types (3D-printer, hairdryer, robotic arm, scanner, ToyCircuit), attributes are provided in this dataset.
For the other four machine types, attributes are concealed.

The \textbf{evaluation dataset} provides the test clips that correspond to the additional training dataset, e.g. data of the same machine types as the additional training dataset. 
Each section consists of 200 test clips, none of which have a condition label (i.e., normal or anomaly) or the domain information. 
No attributes are provided either.

Participants must train a model for a new machine type using only one section per machine type, without hyperparameter tuning using test datasets obtained from the same machine type, and for some of the machine types, without utilizing attribute information.

\subsection{Evaluation metrics} 
\label{sec:metrics} 
We employed the area under the receiver operating characteristic curve (AUC) to evaluate the overall detection performance, while the partial AUC (pAUC) was also utilized to measure performance in a low false-positive rate (FPR) range $\left[ 0, p \right]$.
In this task, we set $p=0.1$.
To evaluate each system under the domain generalization problem setting, we compute the AUC for each domain and pAUC for each section as
\begin{equation}
	{\rm AUC}_{m, n, d} = \frac{1}{N^{-}_{d}N^{+}_{n}} \sum_{i=1}^{N^{-}_{d}} \sum_{j=1}^{N^{+}_{n}}
	\mathcal{H} (\mathcal{A}_{\theta} (x_{j}^{+}) - \mathcal{A}_{\theta} (x_{i}^{-})),
\end{equation}
\begin{equation}
	{\rm pAUC}_{m, n} = \frac{1}{\lfloor p N^{-}_{n} \rfloor N^{+}_{n}} \sum_{i=1}^{\lfloor p N^{-}_{n} \rfloor N^{+}_{n}} \sum_{j=1}^{N^{+}_{n}}
	\mathcal{H} (\mathcal{A}_{\theta} (x_{j}^{+}) - \mathcal{A}_{\theta} (x_{i}^{-})),
\end{equation}
where $m$ represents the index of a machine type,
$n$ represents the index of a section,
$d \in \{ {\rm source}, {\rm target} \}$ represents a domain,
$\lfloor \cdot \rfloor$ is the flooring function,
and $\mathcal{H} (x)$ returns 1 when $x > 0$ and 0 otherwise.
Here, $\{x^{-}_{i}\}_{i=1}^{N^{-}_{d}}$ are the normal test clips in domain $d$ in section $n$ of the machine type $m$ and $\{x_{j}^{+}\}_{j=1}^{N^{+}_{n}}$ are all the anomalous test clips in section $n$ of the machine type $m$.
$N^{-}_{d}, N^{-}_{n}, N^{+}_{n}$ represent the number of normal test clips in domain $d$, the number of normal test clips in section $n$, and the number of anomalous test clips in section $n$, respectively.

The official score $\Omega$ is given by the harmonic mean of the AUC and pAUC scores overall machine types and sections:
\begin{eqnarray}
\Omega &=& h \left\{ {\rm AUC}_{m, n, d}, \ {\rm pAUC}_{m, n} \quad | \quad \right. \nonumber \\
&& \left. m \in \mathcal{M}, \  n \in \mathcal{S}(m), \ d \in \{ {\rm source}, {\rm target} \} \right\},
\end{eqnarray}
where $h\left\{\cdot\right\}$ represents the harmonic mean, $\mathcal{M}$ represents the set of given machine types, and $\mathcal{S}(m)$ represents the set of sections for machine type $m$.
Specifically, $\mathcal{S}(m)=\{00\}$ for the dataset in 2024 Task 2.

\subsection{Baseline systems and results}
\label{sec:baseline}
The task organizers provide a baseline system based on Autoencoders~(AEs), featuring two distinct operating modes. 
This baseline system is the same system employed as the baseline in 2023 Task 2.
Although both modes employ Autoencoder for training, they diverge in the computation of anomaly scores. 
In this paper, we introduce the baseline system along with its detection performance. 
For further information, please refer to \cite{Harada2023}.

\subsubsection{Autoencoder training}
For both operating modes, the AE is trained in the same way.
First, the log-mel-spectrograms of each training sound clips $X = [X_1, \dots, X_T]$ are calculated, where $X_t \in \mathbb{R}^F$ for $t=1,\dots,T$ are the frame-wise feature vectors at frame $t$, $F=128$, is the number of mel-filters and $T$ is the number of time-frames.
For the input of the AE, $P=5$ consecutive frames taken from $X$ are concatenated into one vector as $\psi_t = [X_t^\T, \dots, X_{t + P - 1}^\T]^\T \in \mathbb{R}^{D}$ for each $t$, where $D=P \times F = 640$.
The frame size of short-time Fourier transform (STFT) is 64 ms with the hop size being 50\%. 
The model parameters are trained by optimizing the mean squared error~(MSE) between the system input $\psi_t$ and the reconstructed output $r_\theta (\psi_t)$ for all inputs created from the training data.



\subsubsection{Simple Autoencoder mode}
In this mode, the anomaly score of a sound clip is calculated as the average of the MSE for all input features created from that sound clip, e.g.,
\begin{equation}
A_{\theta}(X) = \frac{1}{DK} \sum_{k = 1}^K \| \psi_k - r_{\theta}(\psi_k) \|_{2}^{2},
\end{equation}
where $K=T-P+1$, and $\| \cdot \|_2$ represents $\ell_2$ norm.

\subsubsection{Selective Mahalanobis mode}
In this mode, the Mahalanobis distance between the system input and reconstructed feature is used to calculate the anomaly score. 
The anomaly score is given as
\begin{align}
&A_{\theta}(X) = \frac{1}{DK} \sum_{k = 1}^K \min\{ D_s (\psi_k, r_{\theta}(\psi_k)), D_t (\psi_k, r_{\theta}(\psi_k))\},
\end{align}
where
\begin{align}
&D_s(\cdot) = \textrm{Mahalanobis}(\psi_k, r_{\theta}(\psi_k), \Sigma_s^{-1}), \\
&D_t(\cdot) = \textrm{Mahalanobis}(\psi_k, r_{\theta}(\psi_k), \Sigma_t^{-1}).
\end{align}
Here, $\Sigma_s^{-1}$ and $\Sigma_t^{-1}$ are the covariance matrices of $r_{\theta}(\psi_k) - \psi_k$ calculated with the source and target domain data of each machine type, respectively.

\subsubsection{Results}
\label{sec:results}

\setlength{\tabcolsep}{1mm} 
\begin{table}[t]
\begin{center}
\caption{Results with Simple Autoencoder mode}
\label{tab:ae_results}
\scriptsize
\begin{tabular}{@{}c c c c p{1pt} c c@{}}
\hline
\ \\[-6.5pt]
Machine type &
Section &
\multicolumn{2}{c}{AUC [\%]} &&
\multicolumn{2}{c}{pAUC [\%]} \\
\cline{3-4} \cline{6-7}
\ \\[-6.5pt]
& & 
\multicolumn{1}{c}{Source} &
\multicolumn{1}{c}{Target} && \\
\hline
\ \\[-6.5pt]
  ToyCar
  &	00 & $66.98 \pm 0.89$ & $33.75 \pm 0.81$ && $48.77 \pm 0.13$\\

  ToyTrain
  &	00 & $76.63 \pm 0.22$ & $46.92 \pm 0.8$ && $47.95 \pm 0.09$\\

  bearing
  &	00 & $62.01 \pm 0.64$ & $61.4 \pm 0.26$ && $57.58 \pm 0.32$\\

  fan
  &	00 & $67.71 \pm 0.7$ & $55.24 \pm 0.91$ && $57.53 \pm 0.19$\\

  gearbox
  &	00 & $70.4 \pm 0.58$ & $69.34 \pm 0.82$ && $55.65 \pm 0.44$\\

  slider
  &	00 & $66.51 \pm 1.66$ & $56.01 \pm 0.29$ && $51.77 \pm 0.35$\\

  valve
  &	00 & $51.07 \pm 0.88$ & $46.25 \pm 1.3$ && $52.42 \pm 0.5$\\
  \hline
\end{tabular}
\end{center}
\end{table}

\setlength{\tabcolsep}{1mm} 
\begin{table}[t]
\begin{center}
\caption{Results with Selective Mahalanobis mode}
\label{tab:mahala_results}
\scriptsize
\begin{tabular}{@{}c c c c p{1pt} c c@{}}
\hline
\ \\[-6.5pt]
Machine type &
Section &
\multicolumn{2}{c}{AUC [\%]} &&
\multicolumn{2}{c}{pAUC [\%]} \\
\cline{3-4} \cline{6-7}
\ \\[-6.5pt]
& & 
\multicolumn{1}{c}{Source} &
\multicolumn{1}{c}{Target} && \\
\hline
\ \\[-6.5pt]
  ToyCar
  &	00 & $63.01 \pm 2.12$ & $37.35 \pm 0.83$ && $51.04 \pm 0.16$\\

  ToyTrain
  &	00 & $61.99 \pm 1.79$ & $39.99 \pm 1.37$ && $48.21 \pm 0.05$\\

  bearing
  &	00 & $54.43 \pm 0.27$ & $51.58 \pm 1.73$ && $58.82 \pm 0.13$\\

  fan
  &	00 & $79.37 \pm 0.44$ & $42.70 \pm 0.26$ && $53.44 \pm 1.03$\\

  gearbox
  &	00 & $81.82 \pm 0.33$ & $74.35 \pm 1.21$ && $55.74 \pm 0.35$\\

  slider
  &	00 & $75.35 \pm 3.02$ & $68.11 \pm 0.63$ && $49.05 \pm 1.0$\\

  valve
  &	00 & $55.69 \pm 1.44$ & $53.61 \pm 0.19$ && $51.26 \pm 0.47$\\
  \hline
\end{tabular}
\end{center}
\end{table}

Tables \ref{tab:ae_results} and \ref{tab:mahala_results} show the AUC and pAUC scores for the two baselines. 
The average and standard deviations of the scores from five independent trials of training and testing are shown in the tables.

\section{Challenge Results}
Challenge results and analysis of the submitted systems will be added for the official submission of the paper to the DCASE 2024 Workshop.

\section{Conclusion}
This paper presented an overview of the task and analysis of the solutions submitted to DCASE 2024 Challenge Task 2. 
We will add challenge results and analysis of the submitted systems for the official submission of the paper to the DCASE 2024 Workshop.

\bibliographystyle{IEEEtran}
\bibliography{refs}

\begin{thebibliography}{10}
\providecommand{\url}[1]{#1}
\def\UrlFont{\rmfamily}
\providecommand{\newblock}{\relax}
\providecommand{\bibinfo}[2]{#2}
\providecommand\BIBentrySTDinterwordspacing{\spaceskip=0pt\relax}
\providecommand\BIBentryALTinterwordstretchfactor{4}
\providecommand\BIBentryALTinterwordspacing{\spaceskip=\fontdimen2\font plus
\BIBentryALTinterwordstretchfactor\fontdimen3\font minus \fontdimen4\font\relax}
\providecommand\BIBforeignlanguage[2]{{%
\expandafter\ifx\csname l@#1\endcsname\relax
\typeout{** WARNING: IEEEtran.bst: No hyphenation pattern has been}%
\typeout{** loaded for the language `#1'. Using the pattern for}%
\typeout{** the default language instead.}%
\else
\language=\csname l@#1\endcsname
\fi
#2}}

\bibitem{koizumi2017neyman}
Y.~Koizumi, S.~Saito, H.~Uematsu, and N.~Harada, ``Optimizing acoustic feature extractor for anomalous sound detection based on {N}eyman-{P}earson lemma,'' in \emph{EUSIPCO}, 2017, pp. 698--702.

\bibitem{kawaguchi2017how}
Y.~Kawaguchi and T.~Endo, ``How can we detect anomalies from subsampled audio signals?'' in \emph{MLSP}, 2017.

\bibitem{koizumi2019neyman}
Y.~Koizumi, S.~Saito, H.~Uematsu, Y.~Kawachi, and N.~Harada, ``Unsupervised detection of anomalous sound based on deep learning and the {N}eyman-{P}earson lemma,'' \emph{IEEE/ACM Transactions on Audio, Speech, and Language Processing}, vol.~27, no.~1, pp. 212--224, Jan. 2019.

\bibitem{kawaguchi2019anomaly}
Y.~Kawaguchi, R.~Tanabe, T.~Endo, K.~Ichige, and K.~Hamada, ``Anomaly detection based on an ensemble of dereverberation and anomalous sound extraction,'' in \emph{ICASSP}, 2019, pp. 865--869.

\bibitem{koizumi2019batch}
Y.~Koizumi, S.~Saito, M.~Yamaguchi, S.~Murata, and N.~Harada, ``Batch uniformization for minimizing maximum anomaly score of {DNN}-based anomaly detection in sounds,'' in \emph{WASPAA}, 2019, pp. 6--10.

\bibitem{suefusa2020anomalous}
K.~Suefusa, T.~Nishida, H.~Purohit, R.~Tanabe, T.~Endo, and Y.~Kawaguchi, ``Anomalous sound detection based on interpolation deep neural network,'' in \emph{ICASSP}, 2020, pp. 271--275.

\bibitem{purohit2020deep}
H.~Purohit, R.~Tanabe, T.~Endo, K.~Suefusa, Y.~Nikaido, and Y.~Kawaguchi, ``Deep autoencoding {GMM}-based unsupervised anomaly detection in acoustic signals and its hyper-parameter optimization,'' in \emph{DCASE Workshop}, 2020, pp. 175--179.

\bibitem{Koizumi2020dcase}
Y.~Koizumi, Y.~Kawaguchi, K.~Imoto, T.~Nakamura, Y.~Nikaido, R.~Tanabe, H.~Purohit, K.~Suefusa, T.~Endo, M.~Yasuda, and N.~Harada, ``Description and discussion on {DCASE}2020 challenge task2: {U}nsupervised anomalous sound detection for machine condition monitoring,'' in \emph{DCASE Workshop}, 2020, pp. 81--85.

\bibitem{Kawaguchi2021}
Y.~Kawaguchi, K.~Imoto, Y.~Koizumi, N.~Harada, D.~Niizumi, K.~Dohi, R.~Tanabe, H.~Purohit, and T.~Endo, ``Description and discussion on {DCASE} 2021 challenge task 2: Unsupervised anomalous detection for machine condition monitoring under domain shifted conditions,'' in \emph{DCASE Workshop}, 2021, pp. 186--190.

\bibitem{Dohi2022}
K.~Dohi, T.~Nishida, H.~Purohit, R.~Tanabe, T.~Endo, M.~Yamamoto, Y.~Nikaido, and Y.~Kawaguchi, ``{MIMII DG}: Sound dataset for malfunctioning industrial machine investigation and inspection for domain generalization task,'' in \emph{DCASE Workshop}, 2022.

\bibitem{Dohi2023DCASE}
K.~Dohi, K.~Imoto, N.~Harada, D.~Niizumi, Y.~Koizumi, T.~Nishida, H.~Purohit, R.~Tanabe, T.~Endo, and Y.~Kawaguchi, ``Description and discussion on {DCASE} 2023 challenge task 2: First-shot unsupervised anomalous sound detection for machine condition monitoring,'' in \emph{DCASE Workshop}, 2023, pp. 31--35.

\bibitem{harada2021toyadmos2}
N.~Harada, D.~Niizumi, D.~Takeuchi, Y.~Ohishi, M.~Yasuda, and S.~Saito, ``{ToyADMOS2}: Another dataset of miniature-machine operating sounds for anomalous sound detection under domain shift conditions,'' in \emph{DCASE Workshop}, 2021, pp. 1--5.

\bibitem{Harada2023}
N.~Harada, N.~Daisuke, T.~Daiki, O.~Yasunori, and Y.~Masahiro, ``First-shot anomaly detection for machine condition monitoring: a domain generalization baseline,'' in \emph{EUSIPCO}, 2023, pp. 191--195.

\end{thebibliography}

\end{sloppy}
\end{document}